\documentclass[aps,preprint,preprintnumbers,nofootinbib,showpacs]{revtex4}
\usepackage{epsfig}
\def\beq{\begin{equation}}
\def\eeq{\end{equation}}
\def\beqa{\begin{eqnarray}}
\def\eeqa{\end{eqnarray}}
\def\lla{\left\langle}
\def\rra{\right\rangle}

\def\ssc{\scriptscriptstyle}
\def\lsim{\mathrel{\raise.3ex\hbox{$<$\kern-.75em\lower1ex\hbox{$\sim$}}} }
\def\gsim{\mathrel{\raise.3ex\hbox{$>$\kern-.75em\lower1ex\hbox{$\sim$}}} }

\begin{document}
\draft
\preprint{{\vbox{\hbox{NCU-HEP-k037}
\hbox{Sep 2010}
%\hbox{rev. Apr 2001}
}}}
%\vspace*{.3in}
%@twocolumnfalse\endcsname

\title{Classical and Quantum Mechanics with Poincar\'e-Snyder Relativity
\vspace*{.3in} }
\author{\bf Otto C. W. Kong and Hung-Yi Lee\vspace*{.2in}}
\email{otto@phy.ncu.edu.tw}
\affiliation{
Department of Physics and Center for Mathematics and Theoretical Physics,
National Central University,~Chung-li,~TAIWAN 32054.
%\footnote{permanent address}  \\
%National Center for Theoretical Science, NTHU, Hsinchu, Taiwan \\
 %\vspace*{.3in}	
}

%\vskip -0.5cm
%\vspace*{2.5in}
\begin{abstract}
The Poincar\'e-Snyder relativity was introduced in an earlier paper of ours as an
extended form of Einstein relativity obtained by appropriate limiting setting of the full
Quantum Relativity. The latter, with fundamental constants $\hbar$ and $G$ built
into the symmetry, is supposed to be the relativity of quantum space-time. Studying
the mechanics of Poincar\'e-Snyder relativity is an important means to get to confront
the great challenge of constructing the dynamics of Quantum Relativity. The mechanics
will also be of interest on its own, plausibly yielding prediction accessible to
experiments. We write the straightforward canonical formulation here, and show that
it yields sensible physics pictures. Besides the free particle case, we also give an
explicit analysis of two particle collision as dictated by the formulation, as well as
the case of a particle rebouncing from an insurmountable potential barrier in the time
direction. The very interesting solution of particle-antiparticle creation and annihilation
as interpreted in a the usual time evolution picture can be obtained, in the simple
classical mechanics setting. We consider that a nontrivial success of the theory, 
giving confidence that the whole background approach is sensible and plausibly on 
the right track. We also sketch the quantum mechanics formulation, direct from
the familiar canoincal quantization, matching to a relativity group geometric
quantization formulation in a previous publication. 
\end{abstract}

%\date{\today}
\maketitle

\section{Introduction}
Poincar\'e-Snyder Relativity is a relativity we introduced recently \cite{036}
as a relativity in between the Galilean or Einstein relativities and the
Quantum Relativity as formulated in Ref.\cite{030}. The idea that Einstein
relativity has to be modified to admit an invariant quantum scale was
discussed by Snyder in 1947\cite{S}. Since the turn of the century, there has
been quite some interest in topic \cite{dsr}. The Quantum Relativity as presented
in Ref.\cite{030} has three basic features: 1/ incorporation of fundamental
invariant through Lie algebra stabilization \cite{CO}, 2/ linear realization
of new relativity symmetry \cite{023}, 3/ the quantum nature to be built in
through two invariants --- essentially independent Planck momentum and Planck
length \cite{030,031}.  Lorentz or Poincar\'e symmetry (of Einstein
relativity) can be considered exactly a result of the stabilization of the
Galilean relativity symmetry. The linear realization scheme in that setting is
nothing other than the Minkowski space-time picture. Ref.\cite{030} arrived
at $SO(2,4)$ as the symmetry for Quantum Relativity or the relativity symmetry
for the `quantum space-time' to be realized as a classical geometry of
a space-like `AdS$_5$'.  Such a mathematically conservative approach leads to
a very radical physics perspective. The `time' of Minkowski space-time is not
just an extra spatial dimension. Its nature is dictated, from the symmetry
stabilization perspective, by the physics of having the invariant speed of light $c$.
The other two new coordinates in our `quantum space-time' picture are likewise
dictated. They are neither space nor time \cite{023,030}. The most important
task at hand is then to understand the nature of the new coordinates and their
roles in physics. It is with the latter in mind that we introduced the
Poincar\'e-Snyder Relativity in Ref.\cite{036}.

In this paper, we want to write down the basics of the canonically formulated
mechanics under the Poincar\'e-Snyder Relativity. The relativity has the symmetry
of $G(1,3)$, the mathematical structure of which is like a `Galilei group' in
$1+3$ (space-time) dimensions \cite{G}. Table~1 shows the mathematical relation
among the various relativity symmetries. Note that the Poincar\'e-Snyder Relativity
is just the Einstein Relativity extended with a new set of transformations
--- the momentum boosts, without involving the quantum invariants.
Poincar\'e-Snyder mechanics is hence expected to have a mathematical formulation
very similar to that under the Galilean or Einstein frameworks. The Poincar\'e-Snyder
Relativity is still a relativity on the 4D Minkowski space-time. However, its
mechanics is to be parameterized by the new independent variable $\sigma$, the
mathematical analog of the time $t$ in the Galilean setting. As the physics
meaning of the variable, with physical dimension of $\frac{time}{mass}$, generally
differs from time, we will use terms lime  $\sigma$-mechanics and $\sigma$-evolution
to mark that. It should be taken as a caution sign against standard dynamical
interpretations. The momentum boosts are like translations by $p^\mu \sigma$, where
the momentum $p^\mu$ has been freed from being the standard mass times velocity
\cite{023} valid in the Galilean and Einstein frameworks.
%%%%%
\footnote{Although Ref.\cite{023} discusses a candidate relativity symmetry different
from the $SO(2,4)$ of Ref.\cite{036} which we follow here, the $ISO(2,4)$ part with the
momentum boosts we focus on here is common to both. The feature is retained in the
contraction $G(1,3)$.}.
%%%%%%
Note that the Einstein rest mass as the magnitude for the energy-momentum four-vector
is not an invariant under the momentum boost transformations.

%\newpage
%%%%%%%%%%%%%%%%%%%%%%%%%%%%%%%%%%%%%%%%%%%%%%%%%%%%%%%%%%%%%%%%%
\begin{table}[t]
 \caption{\footnotesize The various relativities -- matching the generators :
The table matches out the generators for the various relativity symmetries
from a pure mathematical point of view. Note as algebras, the mathematical
structures of translations (denoted by $P_{.}$) or the boosts (denoted by
$K_{.}$ and   $K_{.}'$ -- the so-called Lorentz boosts not included as they are
really space-time rotations) in relation to rotations $J_{..}$ are the same.
Algebraically, translation and boost generators are distinguished only by
the commutation with the Hamiltonian ($H$ and   $H'$). Successive contractions
retrieve $G(1,3)$ and $ISO(1,4)$ from $SO(2,4)$, similar to the more familiar
$G(3)$ and $ISO(1,3)$	
from $SO(1,4)$. In the physics picture under discussion, however, $SO(1,4)$
part of our so-called Snyder relativity $ISO(1,4)$ is {\it different} from
the usual de-Sitter $SO(1,4)$ contracting to $ISO(1,3)$. We consider simply
keeping only the  $P_{\mu}$ and  $J_{\mu\nu}$ generators to reduce from
our Poincar'e-Snyder $G(1,3)$ to the Einstein $ISO(1,3)$. More details in
Ref.\cite{036}.}
%\footnotesize
\begin{center}
\begin{tabular}{|c||c|c|c|c|c|}    \hline\hline
Relativity	& Quantum	& Snyder	& Poincar\'e-Snyder	& Einstein		& Galilean \\
Symmetry	& $SO(2,4)$ & $ISO(1,4)$	& $G(1,3)$		& $ISO(1,3)$	&  $G(3)$\\
\hline
Arena		&`AdS$_5$'	& $M^5$	& $M^4$ (with $\sigma$)	& $M^4$	& $I\!\!R^3$ (with $t$)\\
\hline
\hline
		& $J_{ij}$  	& $J_{ij}$	& $J_{ij}$  	& $J_{ij}$	& $J_{ij}$\\
SO(1,4)	& $J_{i\ssc 0}$		& $J_{i\ssc 0}$	& $J_{i\ssc 0}$        & $J_{i\ssc 0}$	& $K_{i}$ \\
part		& $J_{\ssc 40}$	& $J_{\ssc 40}$	& $K_{\ssc 0}^{\prime}$	& $P_{\ssc 0}$	&  $H$ \\
		& $J_{{\ssc 4}i}$	& $J_{{\ssc 4}i}$	& $K_i^{\prime}$		& $P_i$		&  $P_i$ \\
\hline
		& $J_{\ssc 50}$	&  $P_{\ssc 0}$	&  $P_{\ssc 0}$	& 	&\\	
		& $J_{{\ssc 5}i}$  	&  $P_i$		&  $P_i$		&   	&\\
		& $J_{\ssc 54}$	&  $P_{\ssc 4}$ 	&  $H^{\prime}$	&	&\\
\hline\hline
\end{tabular}
\vspace*{.4in}
\hrule{}
%\hline
\end{center}
\end{table}
%%%%%%%%%%%%%%%%%%%%%%%%%%%%%%%%%%%%%%%%%%%%%%%%%%%%%%%%%%%%%%%%%
When restricted to the setting of Einstein relativity,  Poincar\'e-Snyder
mechanics gives a natural covariant formulation for Einstein mechanics. The same
situation holds after quantization. In fact, as the $G(1,3)$ symmetry admits a
nontrivial $U(1)$ central extension, it gives a better framework for
quantization\cite{036}. We also address some aspects of quantization for
Poincar\'e-Snyder mechanics in this paper.

Explicitly, the algebra of Poincar\'e-Snyder symmetry is given by
\beqa
[J_{\mu\nu}, J_{\lambda\rho}] &=& - %i\hbar\,
( \eta_{\nu\lambda} J_{\mu\rho} - \eta_{\mu\lambda} J_{\nu\rho}
+ \eta_{\mu\rho} J_{\nu\lambda} -\eta_{\nu\rho} J_{\mu\lambda}) \;,
%\eeqa
\nonumber \\ &&
%\beqa
[J_{\mu\nu}, P_\rho ] = - %i\hbar\,
 (\eta_{\nu\rho} P_\mu
- \eta_{\mu\rho} P_\nu ) \;,
\nonumber \\ &&
[J_{\mu\nu}, K_\rho^{\prime}] = - %i\hbar\,
 (\eta_{\nu\rho} K_\mu^{\prime}
- \eta_{\mu\rho} K_\nu^{\prime} ) \;,
\nonumber \\ &&
[K_\mu^{\prime}, H^{\prime}] =  %i\hbar\,
P_\mu \;, % [K_i, H] = P_i
\label{alg}
\eeqa
where the unlisted commutators are all zero and $\eta_{\mu\nu} =( -1, +1, +1, +1)$
\footnote{Note that we follow the metric sign convention adopted
in Ref.\cite{036} which is different from that of Refs.\cite{023,030,031}.}.
To the above, we add the commutator
\beq \label{u1}
[ K_\mu^\prime , P_\nu ]= \eta_{\mu\nu} F \;,
\eeq
characterizing the $U(1)$ central extension needed for quantum mechanics \cite{036,gq}.
The mathematical symmetry had actually been used as a trick to implement quantization
from the latter perspective \cite{pkg}. Since publishing Ref.\cite{036}, we have also
come to realized that a relativity symmetry of essentially the same mathematical
structure has actually been suggested before. Aghassi, Roman, and Santilli
in 1970 argued that (Einstein) relativistic quantum mechanics would better be formulated
with an extended relativity/dynamic symmetry duplicating the structure of the Galilei group
\cite{ARS}. That is the $G(1,3)$ group, which we arrived at from the Quantum Relativity
perspective. Though the physics picture of the extra symmetry transformations is different,
it is interesting to see that in a sense the top-down approach and
the bottom-up approach arise at the same conclusion. We want to emphasize that our basic
perspective gives the radical physics picture of being a non-space-time coordinate. The
feature set the Poincar\'e-Snyder physics we try to formulate apart from all earlier
attempts based on similar mathematics. Moreover, the identification of the extra
transformations as momentum boosts gives
\beq \label{pdef}
p^\mu := \frac{dx^\mu}{d\sigma} \;,
\eeq
which is not to be expected from the other physics frameworks. In fact, the equation should
be taken as a defining one \cite{030,023}, enforcing the compatibility of which with the
standard canonical formulation has nontrivial implications. Note further that $\sigma$
should not be simply taken as Einstein proper time divides by particle rest mass either.
While the latter should hold at the Einstein limit, it has to be relaxed to take
the physics of the Poincar\'e-Snyder Relativity or the Quantum Relativity seriously.
In fact, at the Quantum Relativity or Snyder Relativity level, $\sigma$ as an extra
geometric dimension to space-time has a space-like, rather than time-like, signature.
%%%
\footnote{Though Ref.\cite{ARS} starts by formulating the $G(1,3)$ as a new `dynamical'
symmetry, it assumes the `evolution' parameter $\sigma$ (as we discuss in the section
below) as essentially a time parameter. That is equivalent to keeping momentum as mass
times velocity hence forcing the otherwise independent coordinate/parameter $\sigma$ to
be essentially the Einstein proper time. It actually implies the  $ K_i^\prime$
generators are $not$ independent of the Lorentz boosts  $J_{i\ssc 0}$. Hence, the
role of the $G(1,3)$ symmetry in physics within the framework is really not
much different from that of Ref.\cite{gq}. In our approach, $G(1,3)$ is actually
a basic kinematic symmetry with an independent $\sigma$ parameter playing a
fundamental role.  We will discuss the issue further after we
elaborate our formulation of the $\sigma$-mechanics.}.
%%%%

As an arena for the realization of the relativity symmetry, the Minkowski space-time
supplemented by the (absolute) external parameter $\sigma$ has the transformation
(in the passive point of view)
\beq
x'^\mu = \Lambda^\mu\!_\nu \, x^\nu - p^\mu \sigma - A^\mu
\eeq
and $\sigma$ translation $\sigma' =\sigma - b$. The symmetry action is characterized
by the following realization of the generators :
\beqa
&& J_{\mu\!\nu} \rightarrow -\left( x_\mu \frac{\partial }{\partial x^\nu}
- x_\nu \frac{\partial }{\partial x^\mu} \right) \;,
\qquad
K'_\mu \rightarrow  -\sigma \frac{\partial }{\partial x^\mu}  \;,
\nonumber \\
&& P_\mu  \rightarrow  -\frac{\partial }{\partial x^\mu} \;,
\qquad\qquad
H' \rightarrow  -\frac{\partial }{\partial \sigma} \;.
%\nonumber
\eeqa
%%%%%%%%%%%%%%%%%%%
%{\bf[
%We take the a transformation as given by $\exp(+ \omega^a G_{\! a})$ for parameters $\omega^a$
%and generators $G_{\! a}$, for both active and passive interpretations. One has a passive
%transformation or a transformation of the coordinate frame generated by $G_{\! a}$ by an
%amount $+\omega^a$, not $-\omega^a$. All commutators among the set of generators for the active
%and those for the passive pictures respectively then differ by an overall sign.
%For examples : $\exp(A P_y)(y) = (1+ A P_y + \cdots) (y)$ requires $P_y(y)=-1$, hence
%$P_y$ realized by $\partial_y$; similarly,
%$\exp(\theta J_{\ssc 2\!3})(y) = (1+ \theta J_{\ssc 2\!3} + \cdots) (y)$ requires
%$J_{\ssc 2\!3}(y)= z$, hence $J_{\ssc 2\!3}$ realized by
%$ -(y \partial_z - z \partial_y)$. Note that $J_{\ssc 2\!3}$ rotations of the $yz$-plane
%from the $y$-axis towards the $z$-axis for an angle $\theta$, hence $dy =  z d\theta$.
%The matrix form for $J_{\ssc 2\!3}$ then has the $23$-entry being $+1$. Note that the
%corresponding matrix representation actually gives the whole algebra with a relative
%negative sign to the differential realization.
%]}\\
%%%%%%%%%%%%%%%%%%%
Similar to the Galilean case, a canonical realization \cite{pp} is admissible on
an (one particle) phase space with the introduction of the  momentum variables
conjugate to $x^\mu$. The latter are just $p_\mu$. We illustrate the explicit canonical
formulation here, demonstrating in particular the consistency with the `new' momentum
definition of Eq.(\ref{pdef}).

As discussed above and in Ref.\cite{036}, a straightforward canonical Lagrangian/Hamiltonian
formulation for Poincar\'e-Snyder mechanics looks admissible and sensible, at least in
the mathematical sense. One has to be more cautious with the physics interpretation.
In this paper, we write down the formulation explicitly and looks at a couple of
simple case, namely free particle dynamics and the collisional dynamics of two
otherwise free particles. We take caution against committing to any definite physics
interpretation of the $\sigma$ parameter and the new momentum boost transformations,
including any possible practical implementation of the latter. With the minimalist
approach, we show the results can be sensibly interpreted in a usual time
evolution picture. We focus on classical mechanics, elaborated in the next section.
Quantization of the classical mechanics will then be addressed in sec.III, after which
we conclude in the last section.

\section{Canonical Formulation}
The mathematics of the canonical formulation of the $\sigma$-mechanics is straightforward. 
We mostly want to write down things explicitly for future usage, and to elaborate, whenever
possible, on the physics picture that set it apart from system with similar mathematics.
The most important point to note here is that time $t$, or $x^{\!\ssc 0}=ct$ to be exact,
is now put on the same footing as the spacial position $x^i$, while we have a new absolute
`evolution' parameter $\sigma$. However, $\sigma$ only parameterizes the `evolution' of the
system is a formal sense. It is generally speaking not a sort of time parameter. We 
highlight the feature by terms like $\sigma$-evolution and $\sigma$-Lagrangian.

\subsection{\boldmath $\sigma$-Lagrangian and $\sigma$-Hamiltonian}
We start with a variational principle on the $\sigma$-Lagrangian, for the simple
`one particle' system with configuration variables $x^\mu$ as a prototype and 
specific case of interest. The $\sigma$-Lagrangian would be a function
$\mathcal{L}(x^\mu, \dot{x}^\mu, \sigma)$, where the `dot' here denotes
differentiation with respect to the `evolution' parameter $\sigma$, {\it i.e.}
$\dot{x}^\mu \equiv \frac{dx^\mu}{d\sigma}$.
However, that is just $p^\mu$ [cf. Eq.(\ref{pdef})]; hence
\beq
\mathcal{L}(x^\mu, \dot{x}^\mu, \sigma) =\mathcal{L}(x^\mu, p^\mu, \sigma)\;.
\eeq
The above is specific, only for the case with $x^\mu$ as configuration variables.
The standard procedure introduces the canonical momentum through
$p_\mu= \frac{\partial \mathcal{L}(x^\mu, \dot{x}^\mu, \sigma)}{\partial  \dot{x}^\mu}$,
which can now be written as
\beq
p_\mu= \frac{\partial \mathcal{L}(x^\mu, p^\mu, \sigma)}{\partial  p^\mu} \;.
\eeq
A Legendre transformation gives the  $\sigma$-Hamiltonian as
\beq
\mathcal{H}(x^{\mu},p_{\mu}, \sigma)
= p_{\mu}  \dot{x}^\mu - \mathcal{L}(x^\mu, \dot{x}^\mu, \sigma)
= p_{\mu} p^\mu -\mathcal{L}(x^\mu, p^\mu, \sigma) \;.
\eeq
The action functional on the phase space $\{x^\mu, p_\mu\}$ can be written as
$S[\gamma]\equiv\int_{\sigma^{'}}^{\sigma{''}}d\sigma
\left[ p_{\mu}\dot{x}^{\mu}-\mathcal{H}(x^{\mu},p_{\mu},\sigma) \right]$,
where $\gamma$ denote a phase space trajectory parameterized by $\sigma$.
%: \(\sigma\mapsto x(\sigma),\sigma\mapsto p(\sigma)\) with \(\sigma\in[\sigma{'},\sigma{''}]\).
Variations with respect to $x^\mu$ and $p_\mu$ give us:
\begin{equation}
\delta S[\gamma]
%&=\int_{\sigma^{'}}^{\sigma{''}}d\sigma[\delta p_{\mu}\dot{x}^{\mu}+p_{\mu}\delta\dot{x}^{\mu}-\delta\mathcal{H}(x^{\mu},p_{\mu},\sigma)]\\&
=\int_{\sigma^{'}}^{\sigma{''}} d\sigma \left[ \frac{d(\delta x^{\mu}p_{\mu})}{d\sigma }
+\delta p_{\mu}\dot{x}^{\mu} -\delta x^{\mu}\frac{dp_{\mu}}{d\sigma}
-\frac{\partial \mathcal{H}(x^{\mu},p_{\mu},\sigma)}{\partial x^{\mu}}\delta x^{\mu}
-\frac{\partial \mathcal{H}(x^{\mu},p_{\mu},\sigma)}{\partial p_{\mu}}\delta p_{\mu}
\right] \;.
\end{equation}
Hence, the principle of stationary action yields the Hamiltonian equation of motion
%we have the Euler-Lagrange equation of motion
%\beq
%\frac{d}{d\sigma}\frac{\partial \mathcal{L}}{\partial \dot{x}^\mu}
%- \frac{\partial \mathcal{L}}{\partial  x^\mu} =
%\frac{d}{d\sigma}\frac{\partial \mathcal{L}}{\partial  p^\mu}
%- \frac{\partial \mathcal{L}}{\partial  x^\mu} =0 \;.
%\eeq
%or in the Hamiltonian form
\beqa
\frac{dx^\mu}{d\sigma}&=&\frac{\partial\mathcal{H}(x^\mu,p_\mu,\sigma)}{\partial p_\mu} \;,
\nonumber \\
\frac{dp_\mu}{d\sigma}&=&-\frac{\partial\mathcal{H}(x^\mu,p_\mu,\sigma)}{\partial x^\mu} \;,
\eeqa \label{hem}
which implies
\beq
\frac{d\mathcal{H}}{d\sigma}=\frac{\partial\mathcal{H}(x^\mu,p_\mu,\sigma)}{\partial \sigma} \;. \label{heh}
\eeq

Of course one expect the formal structure of the Lagrangian and Hamiltonian formulation
to maintain when being applied to current framework. The formulations are not restricted
to a specific relativity, though their application to the Einstein relativistic system
is often written in a constrained form \cite{cE}. For a generic set of canonical variables,
we do not necessarily have a relation between the canonical configuration and momentum
variables as given by Eq.(\ref{pdef}) and the $\sigma$-evolution has only the usual
formal description.

Let us elaborate further on the `one particle' system, which has specific features
that will also apply to system with space-time locations as configuration variables.
Altogether, it is easy to see that our special definition in Eq.(\ref{pdef}) requires
\beqa
\mathcal{L}(x^\mu, p^\mu, \sigma) &=& \frac{1}{2} \eta_{\mu\!\nu} p^{\mu} p^\nu -\Phi(x^\mu, \sigma)\;,
\nonumber \\
\mathcal{H}(x^\mu, p_\mu, \sigma) &=& \frac{1}{2} \eta^{\mu\!\nu} p_{\mu} p_\nu +\Phi(x^\mu, \sigma)\;.
\eeqa
The essentially $\frac{1}{2} p_\mu p^\mu$ term, is to be compared to the
$\frac{1}{2}mv^2=\frac{p^2}{2m}$ kinetic energy term common to the Lagrangian and the
Hamiltonian for the case of Galilean relativity. In the latter case, it is enforced by
the particle momentum definition of $p^i := mv^i$. It is interesting to note that there
is no free parameter in the term $\frac{1}{2} p_\mu p^\mu$. There is no explicit role
in the formulation for any particle property like mass. However, one will see in our
discussion of the free particle, {\it i.e. $\Phi=0$}, case below that the term particle
is still justified in the framework.

\subsection{Symplectic Geometric Description}
The phase space for the one-particle system bears the standard symplectic structure
as the cotangent bundle of the Minkowski space-time. The symplectic 2-form is
$\omega= d p_\mu \wedge dx^\mu$, which obviously generalizes directly to an
$N$-particle system. We adopt the notation $\{q^a, p_a\}$ for a set of
canonical coordinates of a generic Poincar\'e-Snyder system. The Poincar\'e
form on $T^*M\times I\!\!R$, where $M$ is the configuration space coordinated by
$\{q^i\}$, has the standard form $\Lambda=p_a \, dq^a - {\mathcal H} \, d\sigma$
with ${\mathcal H}$ as a $\sigma$-Hamiltonian. The Hamiltonian equations of
motion (\ref{hem}) and (\ref{heh}) are summarized in
\[
\mbox{\boldmath $i_{{X}_{\mathcal H}}$} \; \Omega = -d{\mathcal H} \;,
\]
where $\Omega= d \Lambda= dp_a \wedge dq^a - d{\mathcal H} \wedge d\sigma$;
{\boldmath $X_{\!\mathcal H}$} is the  Hamiltonian vector field.
The action functional $S[\gamma]$ is geometrically $\int_\gamma \Lambda\;$.

For a $\sigma$ independent ${\mathcal H}$, we have
\[
\mbox{\boldmath $i_{{X}_{\mathcal H}}$} \; \omega = -d{\mathcal H} \;,
\]
where $\omega= dp_a \wedge dq^a$.
for the function on the cotangent bundle ${\mathcal H}$ as our $\sigma$-Hamiltonian.
Poisson bracket for functions $f$ and $g$ is given by
\beq
-\omega(\mbox{\boldmath $X_f, X_g$})\equiv \{ f, g \}
= \sum_a \left( \frac{\partial f}{\partial q^a} \frac{\partial g}{\partial p_a}
- \frac{\partial f}{\partial p_a} \frac{\partial g}{\partial q^a}  \right) \;,
\eeq
with {{\boldmath $X_{\!\mathcal H}$ }$ =  \{ \cdot, {\mathcal H} \}$}.
In terms of the Poisson bracket, we have
\beq
\frac{df}{d\sigma} = \{f,{\mathcal H}\} +\frac{\partial f}{\partial\sigma} \;.
\eeq
which summaries the mechanics of $\sigma$-(Hamiltonian) evolution.

Everything given in this sub-section is standard in form.

\subsection{Canonical Transformations and Lorentz Transformations}
There are several equivalent definitions of canonical transformations \cite{cE}, here
we follow the most common one, the so-called generating function approach. After writing
down the general theory of canonical transformation, we present generating functions
correspond to the transformations of Poincar\'e-Snyder Relativity symmetry itself.
The latter is the main focus of our presentation here. We also study
infinitesimal version of such generating functions, whenever they leave the
$\sigma$-Hamiltonian function invariant, which are constant of motions with respect to
$\sigma$-evolution. A direct advantage of studying  canonical transformations in the
context of Poincar\'e-Snyder relativity is that the Lorentz transformations can be easily
incorporated into the framework.

We define transformations from phase space variables $(q^a,p_a)$ to $(Q^a,P_a)$ being
canonical if they preserve the structure of Hamiltonian equation of motion. That is
\beq
\frac{dq^{a}}{d\sigma}
=\frac{\partial\mathcal{H}(q^{a},p_{a},\sigma)}{\partial p_{a}} \;,
\qquad
\frac{dp_{a}}{d\sigma}=-\frac{\partial\mathcal{H}(q^{a},p_{a},\sigma)}{\partial q^{a}}\;,
\qquad
\frac{d\mathcal{H}}{d\sigma}=\frac{\partial\mathcal{H}(q^{a},p_{a},\sigma)}{\partial \sigma }\;,
\eeq
{if and only if}
\beq
\frac{dQ^{a}}{d\sigma}=
\frac{\partial\mathcal{K}(Q^{a},P_{a},\sigma)}{\partial P_{a}} \;,
\qquad
\frac{dP_{a}}{d\sigma}=-\frac{\partial\mathcal{K}(Q^{a},P_{a},\sigma)}{\partial Q^{a}}\;,
\qquad
\frac{d\mathcal{K}}{d\sigma}=\frac{\partial\mathcal{K}(Q^{a},P_{a},\sigma)}{\partial \sigma} \;,
\eeq
where $\mathcal{K}(Q^a,P_a)$ is the same $\sigma$-Hamiltonian function represented in
the new phase space variable $(Q^a,P_a,\sigma)$. Its functional form is in general
different from $\mathcal{H}(q^a,p_a,\sigma)$. To find a systematic way to generate such
transformations, we recall that the Hamiltonian equation of motion is derived from phase
space variational principle $\delta\int[p_a\dot{q}^a-\mathcal{H}(q^a,p_a,\sigma)]d\sigma=0$,
treating $\delta p_a$ and $\delta q^a$ independently. The special form of the integrand
is the essence that leads to the canonical form of equation of motion, therefore in
the $(Q^a,P_a)$ coordinates we must also have
$\delta\int P_a\dot{Q}^a-\mathcal{K}(Q^a,P_a,\sigma)=0$. It is not difficult to see the
condition for above two variations hold is that
\beq
P_a\dot{Q}^a-\mathcal{K}(Q^a,P_a,\sigma)=
p_a\dot{q}^a-\mathcal{H}(q^a,p_a,\sigma)+\frac{dF(q^a,Q^a,p_a,P_a)}{d\sigma}
\label{canocondi} \;,
\eeq
since we require the variations at end points to be zero. The variables of function $F$ is
not all independent, they are related by the transformations
\beq
Q^a=Q^a(q^a,p_a,\sigma) \;,
\qquad \qquad
P_a=P_a(q^a,p_a,\sigma)\;.
\eeq
Only half of them are independent variables. For our convenience here, we take $\{q^a,P_a\}$
as independent variables and $F(q^a,P_a,\sigma)$ of the particular form
\beq
F=F_2(q^a,P_a,\sigma)-Q^a P_a.
\eeq
Substitute this equation into Eq.(\ref{canocondi}),  we obtain the following relations:
\beqa
&&p_a=\frac{\partial F_2}{\partial q^a} (q^a,P_a,\sigma)\;,
\qquad\qquad
Q^a=\frac{\partial F_2}{\partial P_a}(q^a,P_a,\sigma)\;,
\nonumber  \\
&&\mathcal{K}(Q^a,P_a,\sigma)
=\mathcal{H}(q^a,p_a,\sigma)+\frac{\partial F_2}{\partial\sigma}(q^a,P_a,\sigma)\;.
\label{if2}
\eeqa
The equations give an implicit relation between the two sets of canonical variables,
from which one can obtain the explicit transformation $(q^a,p_a) \to (Q^a,P_a)$,
and the inverse transformation, so long as the invertibility condition for set of equations
\[
\det\left(\frac{\partial^2F_2}{\partial P_b\partial q^a}\right)\neq 0
\]
is satisfied.

With the above sketch of formulation of canonical transformations, we are ready to
present generating functions that correspond to the transformations of the
Poincar\'e-Snyder Relativity symmetry. We first identify $(q^a,p_a)$ and $(Q^a,P_a)$ as
$(x^\mu,p_\mu)$ and  $(x'^\mu,p'_\mu)$, respectively.
\\
1./ identity transformation ---\\
The generating function for the identity transformation $x'^\mu=x^\mu$, $p'^\mu=p^\mu$
can  obviously be given by
\beq
F_{id}(x^\mu,p'_\mu)=x^\mu p'_\mu \;.
\eeq
%One can easily verify it by simple partial differentiation,
%\beq
%p_\mu=\frac{\partial F_{id}}{\partial x^\mu}=p'_\mu,\qquad x'^\mu=\frac{\partial F_{id}}%{\partial p'_\mu}=x^\mu.
%\eeq
2./ space-time translations ---\\
A space-time translation characterized by a finite four vector $\mathcal{A^\mu}$ can be
generated by
\beq
F_{\ssc P}(x^\mu,p'_\mu)=x^\mu p'_\mu - p'_\mu \mathcal{A}^\mu \;,
%F_3(x'^\mu,p_\mu)=p_\mu\mathcal{A}^\mu-p_\mu x'^\mu
\eeq
giving explicitly
\beq
p_\mu=\frac{\partial F_{\ssc P}}{\partial x^\mu}=p'_\mu \;,
\qquad\qquad
 x'^\mu=\frac{\partial F_{\ssc P}}{\partial p'_\mu}=x^\mu - \mathcal{A}^\mu \;.
\label{trans}\eeq
The associated transformation of momentum is what it has to be, as the last equation implies
$dx'^\mu=dx^\mu$ giving $p'^\mu=\frac{dx'^\mu}{d\sigma}=\frac{dx^\mu}{d\sigma}=p^\mu$.
Of course $F_{\ssc P}$ reduces to $F_{id}$ when $\mathcal{A}^\mu$=0. \\
%%%%%%%%%%%%%%%%
3./ momentum boosts ---\\
Since momentum boosts in Poincar\'e-Snyder Relativity are characterized by a constant
four  momentum $\mathcal{P}^\mu$ and a particular value of the evolution parameter
$\sigma$, its  generating function now must also be $\sigma$ dependent. The generating
function and its partial derivatives reads
\beqa
&& F_{\ssc K'}(x^\mu,p'_\mu,\sigma)
=(x^\mu - \mathcal{P}^\mu \sigma) \, p'_\mu + \mathcal{P}_\mu x^\mu + f(\sigma) \;,
\quad
 f(\sigma)=-\frac{1}{2}\mathcal{P}_\mu\mathcal{P}^\mu\sigma \;,
\nonumber \\
&& p_\mu=\frac{\partial F_{\ssc K'}}{\partial x^\mu}=p'_\mu + \mathcal{P}_\mu \;,
\qquad\qquad \label{cmb}
 x'^\mu=\frac{\partial F_{\ssc K'}}{\partial p'_\mu}=x^\mu - \mathcal{P}^\mu\sigma \;.
\eeqa
Note that $dx'^\mu=dx^\mu-\mathcal{P}^\mu d\sigma$, hence the consistency of the
transformation for the momentum variables. The generating function reduces to $F_{id}$
under the condition that either $\sigma=0$ or $\mathcal{P^\mu}$=0. Strictly speaking
the function $f(\sigma)$ is arbitrary in the formulation. Note that its choice does
not affect the transformation among the canonical variable as given by the second
line of equations. The choice of $-\frac{1}{2}\mathcal{P}_\mu\mathcal{P}^\mu\sigma$
here is what would leaves the Hamiltonian exactly invariant:
\beq
\mathcal{H}(x'^\mu,p'_\mu)=\frac{1}{2} p_\mu p^\mu
+\frac{d}{d\sigma}\!\!\left(\!-\frac{1}{2}\mathcal{P}_\mu \mathcal{P}^\mu\sigma
- p'_\mu\mathcal{P}^\mu\sigma \! \right)=\mathcal{H}(x^\mu,p_\mu)\;.
\eeq
The momentum boosts differ from the other transformations in the symmetry for as
transformations on the space-time, of $x'^\mu=x^\mu -\mathcal{P}^\mu\sigma$, leaves
the Lagrangian or the Hamiltonian only quasi-invariant. The special feature has
consequences we will discuss in a few places below.\\
%%%%%%%%%%%%%%%%
4./space-time rotations (Lorentz transformations) ---\\
Space-time rotations  includes the special cases of 3 dimensional rotations, pure Lorentz
boosts, and the mixture of them. The generating function and its partial derivatives reads
\beqa
F_{\ssc J}(x^\mu,p'_\mu)=p'_\rho \,{\Lambda\!^\rho}\!_\lambda \, x^\lambda \;,
\quad
p_\mu=\frac{\partial F_{\ssc J}}{\partial x^\mu}=  p'_\rho \, {\Lambda\!^\rho}\!_\mu   \;,
\quad
 x'^\mu=\frac{\partial F_{\ssc J}}{\partial p'_\mu}={\Lambda^\mu}\!_\lambda \, x^\lambda \;.
\eeqa
Again, the transformation rule for momentum obtained is consistent with the one
obtained from its definition, namely
$p'^\mu=\frac{dx'^\mu}{d\sigma}={\Lambda^\mu}\!_\nu\frac{dx^\nu}{d\sigma}
={\Lambda^\mu}\!_\nu p^\nu$.
In particular, the generating function for, explicitly, a spatial rotation
 around the $z$-axis (through angle $\theta$) and that for a pure Lorentz boost along
$x$-axis (by velocity $v$) are given, respectively, by
\beqa
&& F_{\ssc 1\!2}=p'_{\!\ssc 0} x^{\!\ssc 0}
+\cos\!\theta \,(p'_{\!\ssc 1} x^{\!\ssc 1}+p'_{\!\ssc 2} x^{\!\ssc 2})
+ p'_{\!\ssc 3} x^{\!\ssc 3}  - \sin\!\theta \,(x^{\!\ssc 1} p'_{\!\ssc 2}
-x^{\!\ssc 2} p'_{\!\ssc 1}) \;,
\\
&& F_{\ssc 1\!0}=\gamma (p'_{\!\ssc 0} x^{\!\ssc 0} + p'_{\!\ssc 1} x^{\!\ssc 1})
+ p'_{\!\ssc 2} x^{\!\ssc 2} + p'_{\!\ssc 3} x^{\!\ssc 3}
-\gamma\frac{v}{c}(  x^{\!\ssc 0} p'_{\!\ssc 1} + x^{\!\ssc 1} p'_{\!\ssc 0} ) \;,
\eeqa
where $\gamma=1/\sqrt{1-(\frac{v}{c})^2}\;$.

Infinitesimal canonical transformations have generators $G(q^a,P_a,\sigma)$ satisfying
\beq
F_2(q^a,P_a,\sigma)=q^a P_a+\epsilon \, G(q^a,P_a,\sigma) \;,
\eeq
with $\epsilon$ being an infinitesimal parameter. From Eq.(\ref{if2}), we obtain
\beq
\mathcal{H} \left(q^a+\epsilon\frac{\partial G}{\partial p_a},
p_a-\epsilon\frac{\partial G}{\partial q^a},\sigma \right)
=\mathcal{H}(q^a,p_a,\sigma)+\epsilon\frac{\partial G}{\partial\sigma} \;.
\eeq
Hence, to leave the $\sigma$-Hamiltonian invariant, {\it i.e.}
$\mathcal{K}(Q^a,P_a,\sigma)=\mathcal{H}(Q^a,P_a,\sigma)$, it is required that
\beq
\frac{dG}{d\sigma}=\{G,\mathcal{H}\}+\frac{\partial G}{\partial\sigma}=0 \;.
\eeq
The above gives the generators as constants of $\sigma$-evolution. For example, for
the free particle $\sigma$-mechanics with $\mathcal{H}=\frac{1}{2}p^\mu p_\mu$, to be
discussed below, generators for the  relativity symmetry transformations give the
following constants of motion :
\beqa
G_{\!\ssc P_\mu} &=&  -p_\mu 		\qquad \mbox{(momentum)}\;,
\nonumber \\
G_{\!\ssc K'_{\nu}} &=& - p_\mu \sigma + x_\mu  \qquad \mbox{(center of mass)}\;,
\nonumber \\
G_{\!\ssc J_{\mu\nu}} &=&-( x_\mu p_\nu -  x_\nu p_\mu) \qquad \mbox{(angular momentum)}\;.
\eeqa
Of course the $\sigma$-Hamiltonian is itself $G_{\!\ssc H'}$. Note that Poisson
brackets among the generators give a realization of the original algebra extended
with a central charge with the value of unity
\footnote{
The $f(\sigma)$ function is Eq.(\ref{cmb}) is what allows the restoration of
$p_\mu$ from $p'_\mu$ is $G_{\!\ssc K'}$, the only case in which
$p'_\mu \ne p_\mu$ in the infinitesimal limit. The necessity and arbitrariness
is related to the quasi-invariant properties of the momentum boosts. The latter
is connected with the appearance of the `extra' $x_\mu$ in $G_{\!\ssc K'_{\nu}}$
or the Noether current and the modification of the commutator
$[ K_\mu^\prime , P_\nu ]$ with a central extension --- all consequences
of the nontrivial cohomology of the $G(1,3)$ group in exact analog to the
Galilean $G(3)$ \cite{036,gq}.
}, {\it i.e.}
\beqa
&&\{G_{\!\ssc J_{\mu\nu}},G_{\!\ssc J_{\lambda\rho}}\}=
\eta_{\nu\lambda }G_{\!\ssc J_{\mu\rho  }} - \eta_{\mu\!\lambda }G_{\!\ssc J_{\nu\rho }}
+ \eta_{\mu\!\rho   }G_{\!\ssc J_{\nu\lambda}}-\eta_{\nu\!\rho   }G_{\!\ssc J_{\mu\lambda}} \;,
\nonumber\\
&&\{G_{\!\ssc J_{\mu\nu}},G_{\!\ssc P_{\lambda}}\}=
\eta_{\nu\lambda }G_{\!\ssc P_{\mu}} -\eta_{\mu\!\lambda }G_{\!\ssc P_{\nu}}\;,
\qquad
\{G_{\!\ssc J_{\mu\nu}},G_{\!\ssc K'_{\lambda}}\}=
\eta_{\nu\lambda }G_{\!\ssc K'_{\mu}} -\eta_{\mu\!\lambda }G_{\!\ssc K'_{\nu}}\;,
\nonumber\\
&&\{G_{\!\ssc K'_\mu}, G_{\!\ssc P_\nu}\} = - \, \{x_\mu, p_\nu \} = -\,\eta_{\mu\!\nu} \;,
\qquad
\{G_{\!\ssc K'_\mu}, G_{\!\ssc H'}\}= - \, G_{\!\ssc P_\mu} \;,
\nonumber\\
&&\{G_{\!\ssc P_\mu}, G_{\!\ssc P_\nu}\}=\{G_{\!\ssc P_\mu}, G_{\!\ssc H'}\}
=\{G_{\!\ssc J_{\mu\nu}},G_{\!\ssc H'}\}=\{G_{\!\ssc K'_\mu},G_{\!\ssc K'_\nu}\}=0 \;.
\eeqa
[{\it cf.} Eqs. (\ref{alg}) and  (\ref{u1})]. Explicitly, all Poisson brackets
among the generators match to the corresponding commutators with a negative
sign illustrating explicitly the formulation as a canonical realization of the
Poincar\'e-Snyder symmetry; hence the central charge as the value for $F$ generator
in Eq.(\ref{u1}) is +1. Fixing the central charge as unity is a special feature
here compared to the corresponding formulation for the Galilean particle. This will
be discussed in the description of free particle mechanics below.

\subsection{Hamilton-Jacobi Equation}
As usual, generating functions for canonical transformations based on the phase space
action lead to Hamilton-Jacobi equation. Consider $\sigma$-evolution given by
\beqa
Q^a &\equiv& q^a(\sigma) = q^a \left(q^a(0), p_a(0), \sigma_{\ssc 0}, \sigma \right)
\nonumber \\
P^a &\equiv& p_a(\sigma) = p_a \left(q^a(0), p_a(0), \sigma_{\ssc 0}, \sigma \right)
\nonumber \eeqa
where $q^a(0)$ and $p_a(0)$ denote the `initial' value of $q^a$ and $p_a$ at
$\sigma=\sigma_{\ssc 0}$. Below within the subsection, we use simply $q^a$ and $p_a$
for the latter. For fixed $Q^a$ and $P_a$, constant in $\sigma$, the generating function
$S \equiv F_2(q^a, P_a, \sigma)$ satisfies
\[
\mathcal{K}(Q^a, P_a, \sigma)
=\mathcal{H}(q^a, p_a, \sigma) + \frac{\partial S}{\partial \sigma} =0 \;.
\]
We have
\[
p_a = \frac{\partial S}{\partial q^a}
\qquad \mbox{and} \qquad
Q^a = \frac{\partial S}{\partial P_a} \;,
\]
hence
\beq
\mathcal{H}\left( q^a, \frac{\partial S}{\partial q^a}, \sigma \right)
+ \frac{\partial S}{\partial \sigma}(q^a, P_a, \sigma) =0 \;.
\eeq
That is the Hamilton-Jacobi equation for $\sigma$-mechanics.

The above is also standard and straightforward, and again
\beq
\frac{dS}{d\sigma}
= \frac{\partial S}{\partial q^a} \dot{q}^a+\frac{\partial S}{\partial \sigma}
= p_a \dot{q}^a - \mathcal{H} \;
\eeq
illustrating that $S$ is  the action re-interpreted as the function
$S(q^a, P_a, \sigma)$.

\subsection{Free Particle $\sigma$-mechanics as an Einstein Limit}
For the free particle $\sigma$-Hamiltonian $\mathcal{H}(x^{\mu},p_{\mu})=\frac{1}{2}p_{\mu}p^{\mu}$,
$\sigma$-mechanics is trivial. The equations of motion, as one expects, are
$\frac{dx^{\mu}}{d\sigma}=p^\mu$, $\frac{dp_{\mu}}{d\sigma}=0$,
and $\frac{d\mathcal{H}}{d\sigma}=0$.
Note that the first equation is only a necessary result as it is exactly our definition
for $p^\mu$ [{\it cf.} Eq.(\ref{pdef})], which is an analog for the velocity definition
$v^i=\frac{dx^i}{dt}$ under Galilean relativity. We can write the constant value for
$\mathcal{H}$ as $-\frac{1}{2}m^2c^2$ with $c$ being the speed of light while $m^2$ is here,
we emphasize, just a number to characterize the value of $\mathcal{H}$. The $p^\mu$ definition
then yields $(m d\sigma)^2=-\frac{1}{c^2} dx_\mu dx^\mu=(d\tau)^2$, where $\tau$ is the familiar
Einstein proper time. Most interestingly, the equation implies
$p^\mu = \pm\, m \frac{dx^{\mu}}{d\tau}$. Taking the positive sign, we have retrieved the
Einstein interpretation of our momentum four-vector provided by the identification of
$m=\frac{d\tau}{d\sigma}$ as the rest mass of an Einstein particle, which is feasible
at least as long as the value of $\mathcal{H}$ is negative. Restricting to negative values
for the $\sigma$-Hamiltonian and positive $\frac{d\tau}{d\sigma}$ , free particle
$\sigma$-mechanics is just a reformulation of
Einstein relativistic mechanics for a free particle. However, instead of having the constant
$m$ as a property characterizing the particle/system, in the $\sigma$-mechanics description,
$m$ gives only essentially the value of the $\sigma$-Hamiltonian. Instead of having particles
with different masses, we have only one (free) particle with different $\mathcal{H}$ values.
More important is the fact that the concept of an invariant particle mass then should not
maintain its meaning in the presence of interactions. Another paramount feature to note is
that as Galilean kinetic energy is not invariant under a velocity boost, our
$\frac{1}{2}p_{\mu}p^{\mu}$ and hence $m$ as a `properties' for the free particle
is not invariant under a momentum boost [{\it cf.} Eq.(\ref{cmb})].

There is another way to see the Einstein nature of the free particle $\sigma$-mechanics.
Consider the quantity $\mathcal{H}_c(x^\mu,p_\mu)=\frac{1}{2}p_\mu p^\mu+\frac{1}{2}m^2c^2$
defined to have vanishing value. If we take $\mathcal{H}_c(x^\mu,p_\mu)$ as a Hamiltonian,
we expect it to give the same `mechanics' with $\mathcal{H}(x^{\mu},p_{\mu})$ as the two
differ only by a constant, except that $\mathcal{H}_c(x^\mu,p_\mu)$ is a constrained
Hamiltonian. Applying the variational principle to $\mathcal{H}_c(x^\mu,p_\mu)$ with the
constraint without reference to $\sigma$ is exactly what is done in the covariant extended
Hamiltonian formulation of Einstein relativistic mechanics \cite{cE}. The standard approach
gives the equations of motion with an undetermined Lagrange multiplier $\lambda$, namely
$\frac{dx^{\mu}}{d\beta}=\lambda p^\mu$ where $\beta$ is a parameter introduced to parameterize
the evolution. The identification $d\sigma = \lambda \, d\beta$ restores results
of our $\sigma$-mechanics. The constrained formulation requires further $\frac{dt}{d\beta}>0$,
for the normal particle, which gives $p_{\!\ssc 0}=-\sqrt{p_ip^i+m^2c^2}$. Note that
from $p_\mu p^\mu = -m^2c^2<0$, we do need $-p_{\!\ssc 0}=p^{\!\ssc 0}>0$ to have
$p^{\!\ssc 0}=\gamma \,mc>0$ for the normal particle, {\it i.e.} positive energy.

Naively, the $p^\mu = -\, m \frac{dx^{\mu}}{d\tau}$, or equivalently $m=-\frac{d\tau}{d\sigma}$,
case differs only by a conventional redefinition of ${\sigma}$ as $-{\sigma}$,
hence unimportant. However, as ${\sigma}$ here is an `absolute' parameter in analog to
Galilean/Newtonian time, we cannot avoid having to deal with a relative negative sign
the $\frac{dx^{\ssc 0}}{d\tau}$ values of two non-interacting `particles'. Hence, a
`particle' can have negative value of $m=\frac{d\tau}{d\sigma}$, which means its
(proper) time `evolves' in $\sigma$ backwards or in opposite direction to the familiar
Einstein particle. It is an antiparticle, essentially in same sense as the
St\"uckelberg-Feynman idea \cite{SF} formulated, in our opinion, in a better conceptual
framework. The antiparticle is the forward time evolving interpretation of a free
particle (in $\sigma$-mechanics) that evolves/moves backwards in time, for as traced
by the absolute/common $\sigma$, its (proper) time `evolves' in opposite direction to that
of a (normal) forward time evolving particle.
When one goes beyond free particle/antiparticle $\sigma$-mechanics, it is easy to
contemplate $\frac{d\tau}{d\sigma}$ evolution that flips sign at some point, giving
something similar to the usual picture of pair creation/annihilation. We will discuss
the kind of scenario below in two-particle scattering/collision analysis.

Next we consider a free particle with $\mathcal{H}=0$. It is quite obvious that the
$\sigma$-mechanics describes in that case the free motion of a particle with zero rest
mass. Of course the particle will be moving with the speed of light when observed
from any Lorentz frame; mathematically, $\frac{d\tau}{d\sigma}=0$ means there is no
$\sigma$-evolution in its proper time. Remember though the mass value of a particle
depends on the choice of reference frames relative to momentum boosts. And we have
been suspending the question of if and how one can implement a momentum boost physically.

Finally, one has to consider positive values of $\mathcal{H}$. Sticking to
$\mathcal{H}= -\frac{1}{2}m^2c^2$, we have $m^2<0$ and the momentum four-vector is
space-like instead of time-like. In the Einstein framework, it looks like a tachyon.
Naturally, we have tachyonic particle as well as tachyonic antiparticles admissible.
Note that Einstein relativity does not excluded tachyonic mechanics, though there may be
difficulties in thinking about interactions among tachyons and normal particles \cite{book}
\footnote{
Einstein relativity put a clear separation between the normal particle world and
the world of tachyonic particles. However, it is important to note that normal
particles appear tachyonic to the tachyons, while tachyons appear to each other normal.
Velocity is relative, hence also the distinction between normal and tachyonic particles
(see Ref.\cite{book} for discussions).
}.
However, given the problem in describing interacting normal particles (no interaction theorem)
\cite{no-int}, it is hard to see the latter as a short-coming of the idea of tachyons being
physical. It may rather be a limitation of the Einstein framework itself.

We see that free particle $\sigma$-mechanics naturally describes a free Einstein particle
of the normal form, as well as the antiparticle, tachyonic particles, and tachyonic
antiparticles, all on the same footing. The true exciting story is in the potential
to describe interactions among such `particles'. While the mathematics of such a
description would be straightforward, the physical interpretation is expected to take
thinking beyond the familiar conceptual framework, in connection to the physical
understanding of the $\sigma$ parameter itself. To that we want to proceed with
upmost caution. A minimal approach is to use only a relational description. That is
a solution path $x^\mu(\sigma)$ gives at each $\sigma$ value the corresponding values
for $x^i$ and $t$ hence $x^i(t)$ implicitly. In fact, our laboratory picture of the
motion is only to be interpreted as $x^i(t)$, maybe with an extra $\sigma(t)$.
We can focus on the former part first, without committing ourselves to the general
physical meaning and plausible measurement of $\sigma$. A good example of the
interpretation is already illustrated by the antiparticle picture, adopted from
St\"uckelberg-Feynman \cite{SF}. We will have more examples below.

Poincar\'e-Snyder free particle has $m$ value characterizing the magnitude of the
four-momentum and hence the $\sigma$-Hamiltonian. The value is dependent on the choice of
reference frame related by momentum boosts. The corresponding formal $\sigma$ evolution
picture (or simply $\sigma$-picture) is what our canonical formulation here gives directly.
The notions of particle, antiparticle, tachyonic particle, and tachyonic antiparticle,
are notions in the Einstein time evolution picture ($t$-picture) within which one freeze
the option of momentum boosts. The latter is the one we use here to interpret results
from our $\sigma$-mechanics.

%It is also interesting to note that if not for
%Eq.(\ref{pdef}), one could choose to write $\mathcal{H}=\frac{z}{2} p_\mu p^\mu$
%where $z$ would be a nontrivial central charge $z$ for the extended algebra
%[{\it cf.} Eq.(\ref{u1})] similar to the Galilean mass. The

\subsection{Particle(s) with Interactions: A First Look}
We have seen that in $\sigma$-mechanics with nontrivial potential, the magnitude
of the four-momentum is no longer constant. Hence, the idea of a particle as a
point mass loses its validity. Here in this subsection, we adopt an abuse of terminology
and keep the term particle to mean a point objection characterized by a space-time position
$x^\mu$ or rather called an Einstein event at each value of $\sigma$. Within a
self-consistent canonical framework, we start to look into simple systems with particle(s) 
in interaction. Readers should bear in mind that, similar to quantum mechanics, a state
of say a two-particle system is expected to resemble one of two Einstein
particles (including antiparticles and tachyons)
only at the asymptotic limit where the interaction is negligible.

\begin{figure}[t]
\caption{\footnotesize Particle-antiparticle annihilation from the time-evolution
picture given as a Poincar\'e-Snyder particle rebounces from an insurmountable
potential barrier in the time direction. Note that after the rebounce, at 
$\sigma>\sigma_o$, the free motion has negative slope in the $t$ Vs $\sigma$ plot,
hence negative energy. It gives an antiparticle as interpreted
in the Einstein relativity time evolution picture ($t$-picture). $t$-picture evolution
is to be read from the $\sigma$-axis upward, as versus the formal $\sigma$-picture 
evolution going left to right. The single  Poincar\'e-Snyder particle is seen
in the $t$-picture as a pair of particle and antiparticle for $t<t_o$ and nothing
for $t>t_o$; they are annihilated at $t=t_o$. We also give the matching
$x$ Vs $\sigma$ plot, with the $t$-picture reading indicated by the short (black) arrows.
}
\hspace{ 0.0cm}
\vspace{-0.5cm}
\centerline{\epsfig{figure=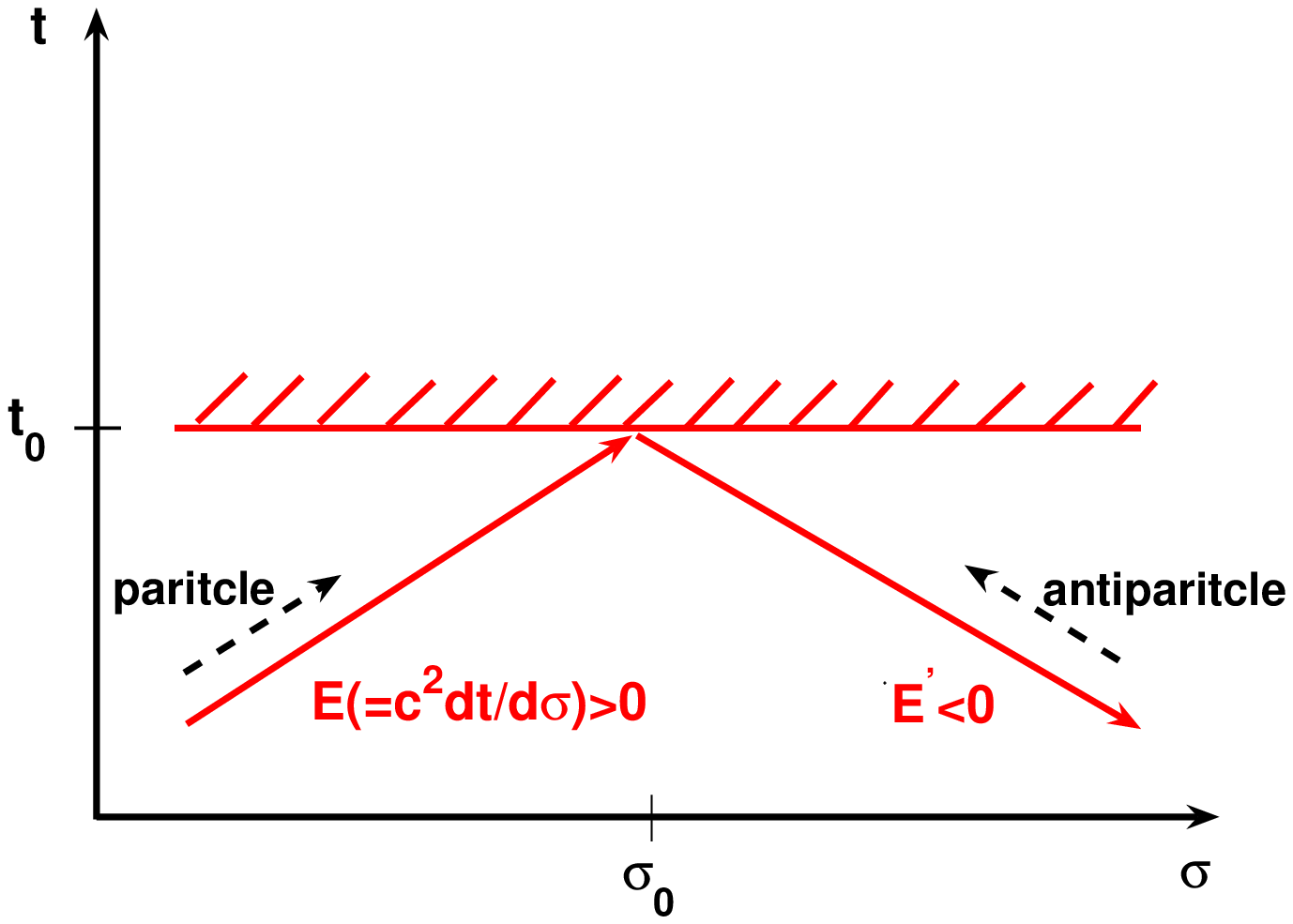,height=10.0cm,width=13.0cm}}\\
\centerline{\epsfig{figure=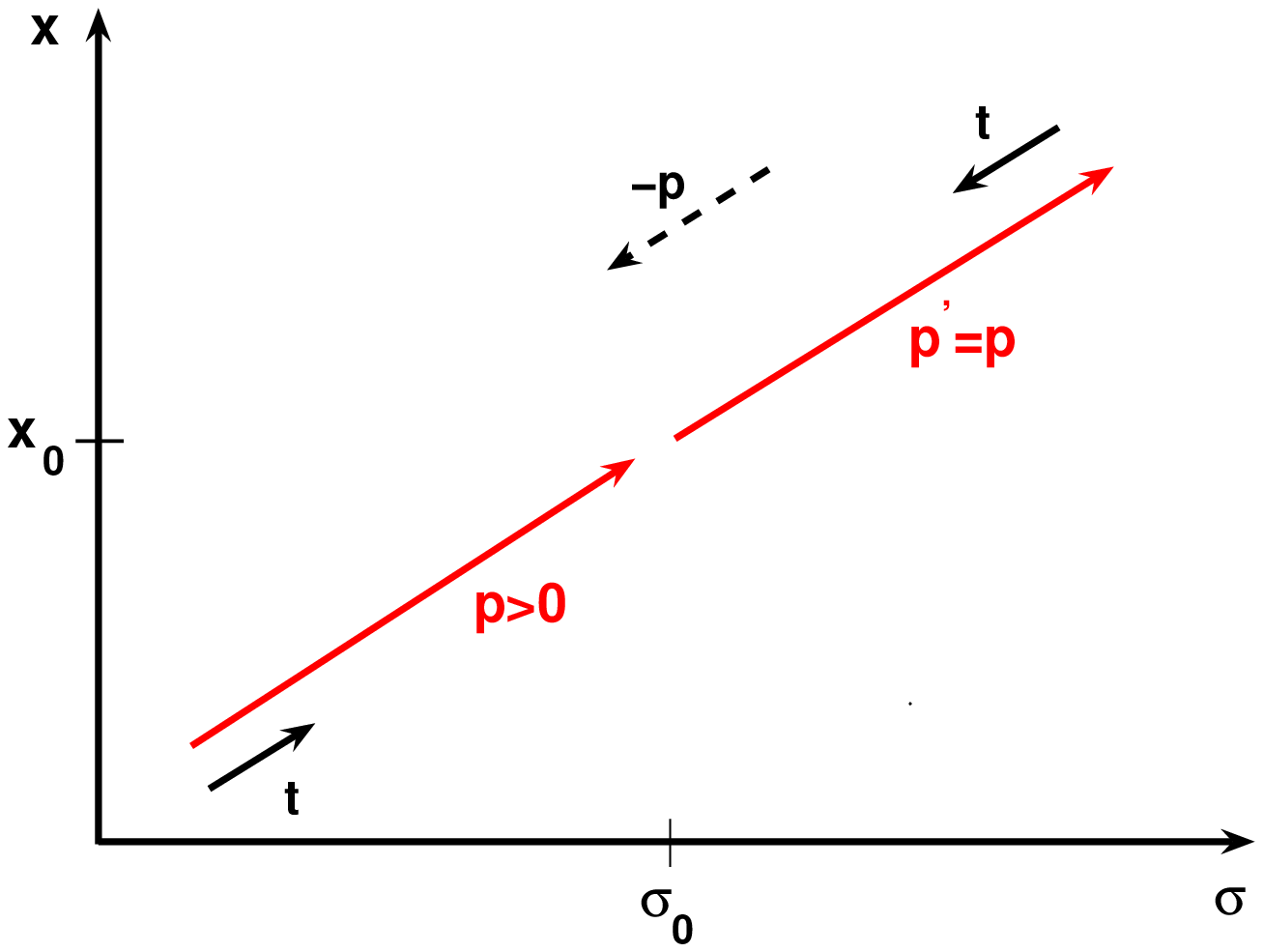,height=10.0cm,width=13.0cm}}
%\centerline{\hline{5in}}
\vspace*{.4in}
\hrule{}
\label{fig1}
\end{figure}

We first illustrate the St\"uckelberg-Feynman picture of particle-antiparticle
annihilation and pair creation. From the above canonical formulation, we have
for a general potential $\Phi(x^\mu)$
\beq \label{pot-int}
\frac{dp_\mu}{d\sigma}= -\frac{\partial\mathcal{H}}{\partial x^\mu}
= -\frac{\partial\Phi}{\partial x^\mu} \;.
\eeq
If one take a potential dependent only on $x^{\ssc 0}$, in fact proportional to
a the delta-function in $x^{\ssc 0}$, the pair annihilation picture will be
resulted. Let us give a bit of details. From the equation, we see that momentum
components in the spacial $x^i$ directions will have constant values. The time
component, however, will not be conserved. So, the energy of an Einstein particle
as essentially given by $\frac{dx^{\ssc 0}}{d\sigma}$ is not constant in $\sigma$.
It change value when $\frac{\partial\Phi}{\partial x^{\ssc 0}}$ is nonzero, hence 
only at one value of $t=x^{\ssc 0}/c$ (say $t=t_o$). As the $\sigma$-Hamiltonian in 
this case still has no explicit dependence on $\sigma$, its value is conserved. 
Hence, the different but constant values of the $\frac{dx^{\ssc 0}}{d\sigma}$ 
before and after the $\sigma$ value with $x^{\ssc 0}(\sigma=\sigma_o)=c \,t_o$ 
have to give the same value for $\frac{1}{2} p^\mu p_\mu$, the value of the free 
particle $\sigma$-Hamiltonian for  $\sigma\ne\sigma_o$. Having 
$p^{\ssc 0}=\frac{dx^{\ssc 0}}{d\sigma}$ flip sign is the only solution. One
can also think about the particle, with initiate $\frac{dx^{\ssc 0}}{d\sigma}>0$,
as confronting an insurmountable potential
barrier in the $t$-direction and bounces back. In the Einstein picture, the sign 
flip changes the particle into an antiparticle, with $p^i$ directions reversed.
When we look at the situation with a time evolution point of view, we get 
particle-antiparticle annihilation at $t=t_o$, as shown in Fig.\ref{fig1}.

A pair creation picture can easily been formulated in a similar fashion. We take
the same potential with $\frac{\partial\Phi}{\partial x^{\ssc 0}}$ nonzero
only at $t=t_o$. All we have to do is to take a $\sigma$-particle with initiate
$\frac{dx^{\ssc 0}}{d\sigma}<0$ `evolving' towards  $\sigma=\sigma_o$ 
corresponding to $t=t_o$. In the $t$-picture, it will be seen as an antiparticle 
emerging from $t=t_o$ with $\sigma$ descreases in time from the value
$\sigma_o$. The solution from the $\sigma$-evolution for $\sigma>\sigma_o$ will give
a $\frac{dx^{\ssc 0}}{d\sigma}>0$ line, hence an Einstein particle also emitted 
at $t=t_o$, completing the pair creation story. Notice that the pair creation and
pair annihilation description here works independent of the magnitude of the
initiate or final momentum four-vector ot the numerical value of $\sigma$-Hamiltonian.
Hence, it works also for massless particles such as a pair of photons (photon
being its own antiparticle). It is easy to imagine that some explicit description
of a realistic potential between two particles at the right initiate condition in 
the $\sigma$-evolution can mediates the pair annihilation and pair creation
parts and describe a realistic experimental setting of such instances. 

Let us now take our analysis to the two particle case.
The $\sigma$-Hamiltonian of two interacting particles of the form
$\mathcal{H}=\frac{1}{2} p_{a}^{\mu} p_{{a}\mu}
+\frac{1}{2} p^{\mu}_{b} p_{{b}\mu} +\Phi(|x_{a}^\mu-x_{b}^\mu|)$
can be canonically transformed into a form with which the associated equations of motion
can be interpreted as a free center motion plus a decoupled one particle motion under
an external influence. The new canonical variables are given by
\beq
X^\mu = \frac{1}{\sqrt{2}}(x_{a}^\mu+x_{b}^\mu) \;,
\qquad\qquad
x^\mu = \frac{1}{\sqrt{2}}(x_{a}^\mu-x_{b}^\mu) \;,
\eeq
and the corresponding $P_\mu$ and $p_\mu$. We have
$\mathcal{H}=\frac{1}{2}P_\mu P^\mu+\frac{1}{2}p_\mu p^\mu+\Phi(\sqrt{2}|x^\mu|)$.
The $(X^\mu, P_\mu)$ set describes free motion of (invariant) mass $M$ given by
$P_\mu P^\mu=-M^2c^2$. The other set $(x^\mu, p_\mu)$ describes a, under the abuse of
terminology, particle motion with $\sigma$-Hamiltonian
$\mathcal{H}_r=\frac{1}{2}p_\mu p^\mu+\Phi(\sqrt{2}|x^\mu|)$,
in which $\Phi(\sqrt{2}|x^\mu|)$ can be interpret as an external potential. In particular
\beq 
\frac{dp_\mu}{d\sigma}= -\frac{\partial\mathcal{H}_r}{\partial x^\mu}
= -\frac{\partial\Phi}{\partial x^\mu} \;.
\eeq
%\begin{eqnarray}
%\frac{dX^{\mu}}{d\sigma}=\frac{\partial\mathcal{H}}{\partial P_\mu}=P^{\,\mu} \\
%\frac{dP_\mu}{d\sigma}=-\frac{\partial\mathcal{H}}{\partial X^\mu}=0\\
%\frac{dx^{\mu}}{d\sigma}=\frac{\partial\mathcal{H}}{\partial p_\mu}=p^\mu\\
%\frac{dp_\mu}{d\sigma}=-\frac{\partial\mathcal{H}}{\partial x^\mu}=-\frac{\partial\Phi}{\partial x^\mu}\\
%\frac{d\mathcal{H}}{d\sigma}=\frac{\partial\mathcal{H}}{\partial\sigma}=0
%\end{eqnarray}
It is obvious that at the limit of vanishing $\Phi$, we do retrieve two free particles.
We also have ${\mathcal{H}_r}$ being constant. However, with nonzero
$\Phi$, $p_\mu p^\mu$ will not be constant and the idea of a constant $m$ being
essentially the magnitude for the momentum $p^\mu$ cannot apply. Nor in the same
sense rest masses for particles $a$ and $b$.

To take the first example of a `nontrivial' interaction, we consider the case of
a collision, {\it i.e.,} an interaction completely localized in space-time. Before
and after the collision, we have two free particles and constant momenta
$p^{\mu}_{a}$  and $p^{\mu}_{b}$ and
$p'^{\mu}_{a}$  and $p'^{\mu}_{b}$, respectively. Notice that our
four-momenta in the problem behave mathematically the same as those of the
three-momenta in Galilean/Newtonian mechanics.
The collisional interaction is supposed to be nontrivial only at the zero
of $x^\mu$. Otherwise, we have $\frac{dp_\mu}{d\sigma}=0$; in fact, we have
$\frac{dp^{\mu}_{a}}{d\sigma}=\frac{dp^{\mu}_{b}}{d\sigma}=0$.
Before collision, at $\sigma <\sigma_{\!o}$, one requires
\begin{figure}[t]
\caption{\footnotesize Collision/scattering of two Poincare\'-Snyder particles 
$a$ and $b$ illustrated, in the center of energy frame.
This is a $t$ Vs $\sigma$ plot, slope of which give the time component of the particle
four-momentum, essentially energy. Negative slope gives a antiparticle as interpreted
in the Einstein relativity time evolution picture ($t$-picture). $t$-picture evolution
is to be read from the $\sigma$-axis upward, as versus the formal $\sigma$-picture 
evolution going left to right. Note that $a$ and $b$ label particle identity in
the $\sigma$-picture, which are not characterized by their masses. And the picture
does not trace the same Einstein particle identity as in the $t$-picture.
}
\hspace{ 0.0cm}
\vspace{-0.5cm}
\centerline{\epsfig{figure=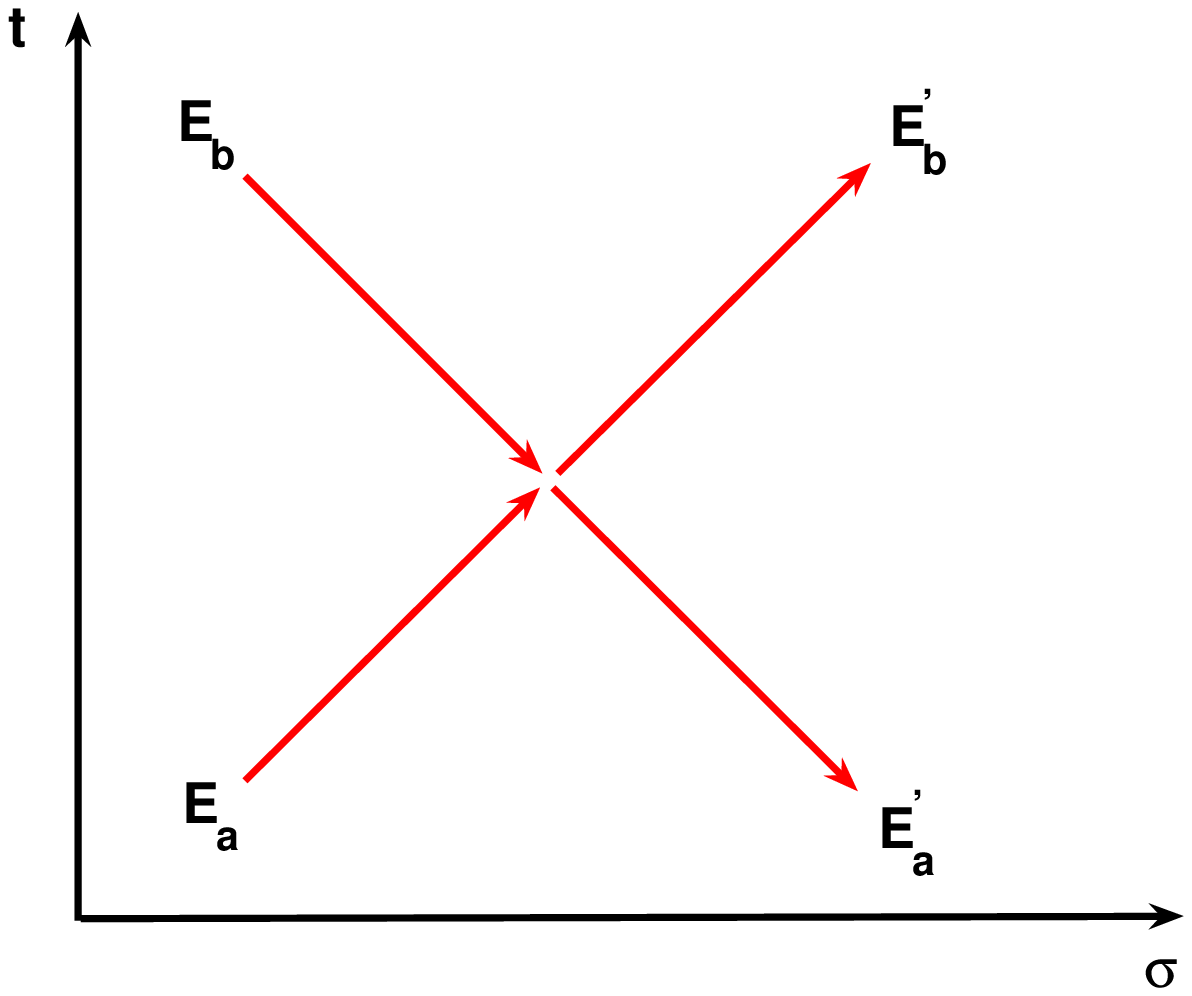,height=10.0cm,width=13.0cm}}\\
%\centerline{\hline{5in}}
\vspace*{.4in}
\hrule{}
\label{fig2}
\end{figure}
\beqa
x_{a}^\mu &=& p^{\mu}_{a} (\sigma -\sigma_{\!o}) + x_{\!o}^\mu \;, \nonumber \\
x_{b}^\mu &=& p^{\mu}_{b} (\sigma -\sigma_{\!o}) + x_{\!o}^\mu \;, \nonumber\\
x^\mu &=& ( p^{\mu}_{a}-p^{\mu}_{b} ) (\sigma -\sigma_{\!o}) \;.
\eeqa
And we have similar picture after the collision. With total four-momentum ($P_\mu$)
conservation, one can have only $(p'^{\mu}_{a}, p'^{\mu}_{b})= (p^{\mu}_{b}, p^{\mu}_{a})$,
as the value of $\sigma$-Hamiltonian is also constant. So we have a two to two
scattering, with in fact the mass values interchanged after the collision
(at $\sigma>\sigma_{\!o}$). At the center of four-momentum frame, in particular, we have
$P_\mu=0$ or $p^{\mu}_{a}=-p^{\mu}_{b}$. The two particles have equal and opposite
four-momenta hence one has to have a negative energy value. Again, one is to be
interpreted as an antiparticle in the $t$-picture.
The story is actually more complicated than that. As illustrated in Fig.\ref{fig2},
tracing the time evolution in the $t$-picture, instead of a particle with
$p^{\mu}_{a}$ colliding/scattering with one with $p^{\mu}_{b}$ to give a complete
four-momentum exchange, we have one with $p^{\mu}_{a}$ (assuming $p^{\!\ssc}_{a}>0$)
and another with $p'^{\mu}_{a}$ but interpreted as an antiparticle having positive
physical energy and four-momentum $-p'^{\mu}_{a}(=-p^{\mu}_{b})$ colliding/scattering
to give a particle with  $p'^{\mu}_{b}=p^{\mu}_{a}$ and a antiparticle with $-p^{\mu}_{b}$.
Scattering of particles $a$ and $b$ in the $\sigma$-picture becomes scattering
of particles $a$ of $\sigma <\sigma_{\!o}$ and $a$ of $\sigma >\sigma_{\!o}$
into particles particles $b$ of $\sigma <\sigma_{\!o}$ and $b$ of $\sigma >\sigma_{\!o}$.
The idea of identity of particles is not the same in the two pictures. At the center of
momentum frame in the $\sigma$-picture, we have the $t$-picture scattering of a particle with
an antiparticle, but with identical four-momenta as $-p'^{\mu}_{a}= -p^{\mu}_{b}=p^{\mu}_{a}$.
Mathematically, eliminating $\sigma$ gives for particle $a$ of $\sigma <\sigma_{\!o}$ and
$\sigma >\sigma_{\!o}$
\[
x^i_a = \frac{c p^i_a}{p^{\ssc 0}_a}(t_{\!a}-t_{\!o}) +  x_{\!o}^i
\quad		\mbox{and}  	\quad
x'^i_a = \frac{c p'^i_a}{p'^{\ssc 0}_a}(t'_{\!a}-t_{\!o}) +  x'^i_{\!o} \;,
\]
with $t_{\!a},t'_{\!a}>t_{\!o}$. At the center of four-momentum frame,
$\frac{c p'^i_a}{p'^{\ssc 0}_a}=\frac{c p^i_a}{p^{\ssc 0}_a}$ as a common negative
sign shows up in both the numerator and denominator. That gives $x^i_a= x'^i_a$ at any
value of time. So, the particle and antiparticle are always at the same position
in space, but have different $\sigma$ values which change in opposite directions.
At time $t_{\!o}$, the $\sigma$ values become the same and they interact and switch
their four-momenta. The particles masses are of course the same as a result of 
the choice of frame of reference. Particle rest mass is reference dependent under
momentum boosts. One can consider however the two particles $a$ and $b$
to have different identities characterized by some conserved charges, like one
being an electron and the other a quark. The $t$-picture story is then essentially 
one of electron-positron annihilation creating a quark-antiquark pair. The 
particle-antiparticle at the same position and two pairs of the same masses
part maybe consdier an artefact of the choice of reference frame. For example,
taking a center of energy frame with $p^i_a= p^i_b$ for the two Poincar\'e-Snyder 
particles will give the $t$-picture of particle-antiparticle moving towards
or away from one another. Notice that a momentum boost frame transformation
transforms also the potential $\Phi$ in general. 

The collision/scattering story looks interesting and quite successful. We have provided,
for instance, a description of something like particle-antiparticle creation and
annihilation  in a classical particle dynamics setting in our $\sigma$-mechanics. 
That is a nontrivial success of the formulation. Further investigations
of $\sigma$-mechanics with nontrivial interactions may provide a lot more in the future.
We note in passing the as the $\sigma$-Hamiltonian for a tachyon is positive, the
$\sigma$-Hamiltonian  for a system of a particle and a tachyon can be zero. It is
possible for such a pair to annihilate into nothing --- a truly intriguing scenario.

Here in this paper, we want to focus mostly on providing and justifying a formulation
of $\sigma$-mechanics --- the direct canonical formulation presented. We will stop
here on analysis of specific $\sigma$-Hamiltonian. Next, we look briefly at the 
corresponding quantum mechanics.

\section{Quantization in Perspective}
Independent of of its role as an intermediate framework between the elusive
Quantum Relativity and the familiar relativities \cite{036},
one argument in favor of replacing the Poincar\'e symmetry by the Poincar\'e-Snyder
$G(1,3)$ is the more natural behavior under quantization with space-time position
operator $X^\mu$ contained in the symmetry algebra \cite{036,ARS}. We have presented
an explicit description of the quantization of free particle $\sigma$-mechanics under
the geometric perspective of the $U(1)$ central extension of $G(1,3)$ \cite{036},
following closely the approach of Refs.\cite{gq,pkg}.
At the classical level, the $\sigma$-Lagrangian ${\mathcal L}=\frac{1}{2} p^\mu p_\mu$
is only quasi-invariant under the momentum boosts
$\dot{x}^\mu \rightarrow \dot{x}^\mu + {\mathcal P}^\mu$ ($\dot{x}^\mu=p^\mu$) :
\beq
{\mathcal L} \rightarrow {\mathcal L}^\prime
= {\mathcal L} + \frac{d}{d\sigma}\eta_{\mu\nu}
\left( \frac{1}{2} {\mathcal P}^\mu {\mathcal P}^\nu \sigma -x^\mu {\mathcal P}^\nu\right)
\equiv {\mathcal L} +\frac{d}{d\sigma} \Delta(\sigma, x^\mu; {\mathcal P}^\mu) \;.
\eeq  % x^\mu actually for the new frame
When compared to the Galilean case, the only difference is the missing of a parameter
corresponds to the central charge (which is the particle mass in the Galilean case).
Note that the $\Delta$ term though having no effect on the equations of motion,
does affect the definition of the Noether charges. Geometrically, the nontrivial
cohomology of the $G(1,3)$ group gives rise to the nontrivial $U(1)$ central extension
which represents the phase transformation in the quantum description of a state.
The feature is missing in the usual Einstein relativistic quantum mechanics with
trivial group cohomology.

On the quantum level, we have the central extension given by
\[
[K_\nu^\prime , P_\mu] = i\hbar \,\eta_{\mu\!\nu} F \;,
\]
where we have re-scaled the algebra to put in the $i\hbar$ for the quantum case
[{\it cf.} Eq.(\ref{u1})]. The generators of the momentum boosts $K_\mu^\prime$
have the properties of covariant space-time position operator, as illustrated in
Ref.\cite{036}. From the analysis of the Quantum Relativity level \cite{030},
we actually expect noncommuting $X_\mu = \frac{1}{\kappa \,c} K_\mu^\prime$
%%.
That perspective reinforces the result that unlike the Galilean
$X_i= \frac{1}{m} K_i$, $K_\nu^\prime$ is essentially the position operator
for all `particle' state independent of any `particle' properties like $m$.
Again, $m$ is not an invariant under the momentum boosts.
The commutator of the central extension is to be identified directly as
the Heisenberg commutation relation. We have hence a consistent result, namely
to arrive at quantum mechanics as we know it, the central charge has to be taken
as unity instead of a free parameter.
The other commutator relation re-casted as
\beq
P_\mu = \frac{d K_\mu^\prime}{d\sigma} = \frac{1}{i\hbar} [K_\mu^\prime, H^\prime]
\eeq
gives the Heisenberg equation of motion.

Next, we write down basic results from the more common quantization framework, namely
a canonical quantization based on the Schr\"odinger picture.

The Schr\"odinger equation for $\sigma$-mechanics, or called the
$\sigma$-dependent covariant Schr\"odinger equation,
\beq \label{se}
i\hbar\frac{d}{d\sigma} \left|\,{\psi,\sigma}\rra=\hat{\mathcal{H}}\left|\,{\psi,\sigma} \rra \;
\eeq
is obtained in Ref.\cite{036} from the group/geometric analysis on the free particle case.
The operator representation for the $\sigma$-Hamiltonian
$\hat{\mathcal{H}} \rightarrow i\hbar\frac{d}{d\sigma}$ on the
Hilbert space of quantum states $\left|{\psi,\sigma}\rra$ is the starting point for
the standard canonical quantization, in accordance with the classical canonical
formalism above. That is,
$\mathcal{H}(x^\mu,p_\mu)\rightarrow\hat{\mathcal{H}}(\hat{x}^\mu,\hat{p}_\mu)$. The operators
$\hat{x}^\mu$ and $\hat{p}_\mu$ satisfy the fundamental canonical commutation relations
\beq
[\hat{x}^\mu,\hat{x}^\nu]=0 \;,
\qquad
[\hat{p}_\mu,\hat{p}_\nu]=0 \;,
\qquad
[\hat{x}^\mu,\hat{p}_\nu]=i\hbar{\delta^\mu}_{\!\nu} \;.
\eeq
At least formally, an $x$-representation can be written by taking
\[
\hat{x}^\mu\rightarrow x^\mu \;,
\qquad
\hat{p}_\mu\rightarrow -i\hbar\frac{\partial}{\partial x^\mu} \;,
\]
giving the generic particle Hamiltonian $\hat{\mathcal{H}}(\hat{x}^\mu,\hat{p}_\mu)
=\frac{-\hbar^2}{2}\,\partial^\mu\partial_\mu+\Phi(x^\mu,\sigma)$.  Explicitly, the
Schr\"odinger equation becomes
\beq
i\hbar\frac{\partial\psi(x^\mu,\sigma)}{\partial\sigma}
    =\left[\frac{-\hbar^2}{2}\,\partial^\mu\partial_\mu+\Phi(x^\mu,\sigma)\right]\psi(x^\mu,\sigma)\;,
\eeq
where $\psi(x^\mu,\sigma)$ is the space-time wavefunction for the state at $\sigma$.
In the case that the `potential' does not depend on $\sigma$ explicitly, {\it i.e.} $\Phi=\Phi(x^\mu)$,
we can perform separation of variables and obtain a generalized Klein-Gordon equation:
\beqa
&& \psi(x^\mu,\sigma)=\phi(x^\mu)\,\Sigma(\sigma)
\nonumber\\
\Rightarrow \qquad &&
i\hbar \, \frac{\partial\Sigma(\sigma)}{\partial\sigma}
%=\frac{-\hbar^2}{2}\frac{1}{\phi(x^\mu)}\partial^\mu\partial_\mu\phi(x^\mu)+\Phi(x^\mu)\frac{1}{\phi(x^\mu)}
=-\frac{1}{2} \, m^2c^2 {\Sigma(\sigma)}   \nonumber\\
\mbox{and} \qquad &&
\partial^\mu\partial_\mu\phi(x^\mu)-\frac{m^2c^2}{\hbar^2} \, \phi(x^\mu)= \frac{2}{\hbar^2}\, \Phi(x^\mu) \;,
\eeqa
where we have, following the analysis  of the classical mechanics, written the eigenvalue
of the $\sigma$-Hamiltonian as $-\frac{1}{2}m^2c^2$.  Of course for $\Phi=0$, one obtains the
Klein-Gordon equation for Einstein relativistic mechanics of a free (spin 0) particle.

Taking the Schr\"odinger equation on an abstract state to that of the $x$-representation
can also be performed in the standard fashion. One assume a complete set of orthonormal
space-time position eigenstates $\hat{x}^\mu \!\left|\,{\bf{x}}\rra = x^\mu \!\left|\,{\bf{x}}\rra$
satisfying
\[
\lla \bf{x'} |\, \bf{x} \rra = \delta^4(\bf{x'}-\bf{x})  \;
\qquad \mbox{and}\qquad
  \int \! d^4{x}\left| \,{\bf{x}}\rra \! \lla \bf{x} \right| = 1 \;. % \ -\infty<x'^\mu<\infty
\]
The wavefunction $\psi(x^\mu,\sigma)$ has values given by the expansion coefficients
$\lla \bf{x}| \, {\psi,\sigma}\rra$ in
\[
\left|\, {\psi,\sigma}\rra = \int\! d^4x \left| \,{\bf{x}}\rra \! \lla \bf{x} |\,{\psi,\sigma}\rra \;.
\]

A question arise as to the interpretation of the wavefunction. A naive Born
interpretation looks like possible and at the same time problematic. It is
easy to see that the integral of wavefunction magnitude-squared can natural
give the relative probability of finding the quantum state to be within
the space-time region. However, the normalization as unit probability
sounds unconventional. Similar comment applies to the the expectation
value of an appropriately defined operator over a particular state. Here,
we would rather if the question open for the moment, awaiting further analyses
of various system Hamiltonians to give more insight for addressing the
question.

We will give a coherent state representation {\it a la} Klauder\cite{K} in
a forth-coming paper\cite{042} in which we also discuss the case of
harmonic oscillator under Poincar\'e-Snyder relativity as well as the
path-integral picture.

\section{conclusions}
We introduced the Poincar\'e-Snyder relativity as an extended version of Einstein
relativity with $G(1,3)$ symmetry \cite{036}. It is a contraction limit of the
full Quantum Relativity. The relativity has in addition to the Poincar\'e $ISO(1,3)$
transformations an extra class of transformations called momentum boosts \cite{023},
dependent on the new parameter $\sigma$. We want to be exceptionally cautious before
committing to a particular physical picture about the parameter and the transformations.
So, we take a minimalist approach here, trying to what could be the
Poincar\'e-Snyder mechanics and if and how we could make sense out of it before
even understanding about the physics of the $\sigma$ parameter and the transformations.

We show here there is a straightforward canonical formulation of classical and
quantum mechanics under the Poincar\'e-Snyder relativity, with $\sigma$ as a formal
`evolution' parameter. The formulation gives free particle mechanics essentially
the same as Einstein relativity though without the rest mass as an intrinsic
defining properties of the particle. And a Poincar\'e-Snyder can be a particle,
antiparticle, tachyon, and tachyonic antiparticle, under the time evolution
($t$-picture) of Einstein relativity. So long as a human observer is concerned
it is clear that it is the $t$-picture that can describe our laboratory experience.
This is the wisdom from St\"uckelberg\cite{SF}. The interpretation is achieved
within a particular reference frame relative to the momentum boosts. As we have
no prior experience or do not know about momentum boosts transformations in
experiments so far, it is in fact natural to expect that we have been studying
physics essentially or approximately in one specific momentum boost frame.

To go beyond the free particle case, we look at the case with an insurmountable 
potential barrier in the time direction. The `rebounce' turns a particle into
an antiparticle. The two particle collision is the next setting, and the first case of 
nontrivial interaction we analyze here. All the results are dictated directly by the 
mathematics within the canonical picture, and looks interesting and quite successful. 
We have provided, for instance, a direct description of something like 
particle-antiparticle creation and annihilation in a classical particle dynamics 
setting in our $\sigma$-mechanics.  That is a nontrivial success of the formulation.

Quantization of the canonical formulation is also straightforward. It matches
exactly to the direct group geometric quantization picture based on the $U(1)$
central extension of $G(1,3)$ \cite{036}. The latter illustrates that $G(1,3)$
has the superior properties for giving rise to the conventional relation
between canonical realization as in classical mechanics and projective
representation as in quantum mechanics --- one the the $ISO(1,3)$ symmetry
fails.  The potential to use our Ponicar\'e-Snyder formulation to re-analyze every
aspects of relativistic quantum mechanics including foundation issues, information
theoretical aspects, and various applications is unlimited. It is likely to
provide, among other things, a better picture to describe macroscopic
superposition of states localized otherwise at different space-time points
\cite{B}.

What we have taken here is an important step towards understanding and building
the dynamics for our Quantum Relativity, and plausibly the necessary step to find
direct connection with the experimental front at the more familiar scale (rather
than Planck scale). The formulation of Poincar\'e-Snyder mechanics here gives
us good confidence that we are on the right track. And the theory is of interest
on its own. It could provides a new perspective and mathematical setting to
handle the usual (Einstein) relativistic mechanics both at the classical and
quantum levels. It would also describe a much boarder class of phenomena to
be studied experimentally.

\bigskip
\bigskip

\noindent{\em Acknowledgements :-\ }
We thank D.-N. Cho for helping to produce the figures.
The work is partially support by  the research grants No.
96-2112-M-008-007-MY3 and No. 99-2112-M-008-003-MY3 from the NSC of Taiwan.

\end{document}